%% file: main.tex
\documentclass[conference]{IEEEtran}
\IEEEoverridecommandlockouts
\usepackage{cite}
\usepackage{url}
\usepackage{booktabs}
\usepackage{graphicx}
\usepackage{tikz}
\usepackage{algorithm}
\usepackage{paralist}
\usepackage{enumitem}
\usepackage{algpseudocode}
\usepackage{multirow}
\usepackage{balance}
\usepackage{float}
\usepackage{array}
\usepackage{amsmath,amssymb,amsfonts}
\usepackage{textcomp}
\usepackage[caption = false]{subfig}
\usepackage[super]{nth}

\usepackage{arydshln}

\def\BibTeX{{\rm B\kern-.05em{\sc i\kern-.025em b}\kern-.08em
    T\kern-.1667em\lower.7ex\hbox{E}\kern-.125emX}}

\newcommand{\RQcomplexity}{Why are faults not detected at testing time?}
\newcommand{\RQcontext}{Which elements of the field are involved in field failures?}
\newcommand{\RQsteps}{How many steps are needed to reproduce a field failure?}
\newcommand{\RQmanifest}{What kinds of field failures can be observed?}

\newcommand{\RQcomplexityN}{RQ1}
\newcommand{\RQcontextN}{RQ2}
\newcommand{\RQmanifestN}{RQ3}
\newcommand{\RQstepsN}{RQ4}

\newcommand{\CHANGED}[1]{#1}
\newenvironment{changed}{\color{black}}{\color{black}}

\newenvironment{pdefinition}[1]
    {\medskip \textbf{#1}:}
    { \smallskip}
    
\newenvironment{finding}[1]
    {\emph{#1}:}
    {}
    
\newenvironment{researchquestion}
    {\smallskip \noindent \begin{em}}
    { \end{em}}

\begin{document}
%
\title{An Exploratory Study of Field Failures}

\author{\IEEEauthorblockN{Luca Gazzola, Leonardo Mariani, Fabrizio Pastore, Mauro Pezz\`{e}}
\IEEEauthorblockA{
Department of Informatics Systems and Communication (DISCo)\\
University of Milano - Bicocca\\
Milano, Viale Sarca 336\\
Email: \{luca.gazzola,mariani,pastore,pezze\}@disco.unimib.it}}

\maketitle

\begin{abstract}
Field failures, that is, failures caused by faults that escape the testing phase leading to failures in the field, are unavoidable. 
Improving verification and validation activities before deployment can identify and timely remove many but not all faults, and 
users \CHANGED{may} still experience a number of annoying problems while using their software systems. 

This paper investigates the nature of field failures, to understand to what extent further improving in-house verification and validation activities can reduce the number of failures in the field, and frames the need of new approaches that operate in the field.
 
We report the results of the analysis of the bug reports of five applications belonging to three different ecosystems, propose a taxonomy of field failures, and discuss the reasons why failures belonging to the identified classes cannot be detected at design time but shall be addressed at runtime. 
We observe that many faults (70\%) are intrinsically hard to detect at design-time


%
\end{abstract}

\begin{IEEEkeywords}
field faults, field failures, field testing, field-intrinsic faults, failure context, software testing
\end{IEEEkeywords}

\input{introduction}

\input{categories}

\input{researchquestions}

\input{subjects}

\input{procedure}
\input{results}

\input{threats}

\input{discussion}

\input{related}

\input{conclusion}

\section*{Acknowledgment}
This work has been partially supported by the H2020 Learn project, which has been funded under the ERC Consolidator Grant 2014 program (ERC Grant Agreement n. 646867) and the GAUSS national research project, which has been funded by the MIUR under the PRIN 2015 program (Contract 2015KWREMX).

\bibliographystyle{IEEEtran}
\bibliography{fieldFailuresStudy}

\end{document}

%% file: introduction.tex
\section{Introduction}

Software field failures are failures that occur in the field with sometimes severe consequences on users and organizations, such as
customer dissatisfaction, economic losses and legal issues. 
\emph{Field failures} are caused by faults that escape the in-house testing activities and are not detected and repaired before the software is released in the field.
We denote such faults as \emph{field faults}. 

Field failures may depend on weak testing activities and poor development practices.  
However, they may also derive from factors that prevent the failures to be detected and the corresponding faults to be removed before the software is released, such as
when the conditions that trigger the failure are impossible to reproduce in the testing environment and when the number of combinations to be executed goes beyond any reasonable limit.

An example of conditions impossible to reproduce in-house is the extraordinary system load that derives from millions of people connected over the Internet to watch an exceptional sport event like the final match of the European Champions League or the US Super Bowl streamings~\cite{superbowlWeb2015}. 
It is impossible to reproduce the same environment conditions to accurately test the system in-house for revealing the uncovered and failure-prone behaviors that may occur in the field, as it happened in 2016 when the CBS app failed to stream the Super Bowl match to several customers~\cite{superbowlAppNotworking}. 

An example of amount of combinations impractical to execute with in-house testing is the extraordinary cardinality of the Microsoft environment configurations~\cite{Murphy-InVivo-ICST-2009} that reaches trillion of combinations of the configuration parameters.

Field faults that cannot be detected with in-house testing approaches might be  more easily addressable in the field where the diversity and complexity of the execution environment could be exploited in the verification activity. 
Field faults have attracted the interest of both academia, mainly in the context of service-based~\cite{Hielscher2008,Sammodi2011} and caching systems~\cite{Murphy-InVivo-ICST-2009}, and industry, with approaches like Netflix that injects faults in the production systems to validate scenarios that are impossible to test in-house~\cite{Basiri:Netflix:ISSRE:2016}.

Studies of field faults have considered many aspects, such as fault distribution~\cite{Hamill-TrendsInFaults-TSE-2009,Fan-NuclearFailures-SF-2013}, fault locality~\cite{Hamill-TrendsInFaults-TSE-2009}, fault locations~\cite{Ostrand-FaultDistribution-ISSTA-2002}, activities and types of human errors that introduce faults~\cite{Leszak-ClassificationOfDefects-JSS-2002}, relations between fault types, failure detection and failure severity~\cite{Hamill-FaultTypesDetectionSeverity-SQJ-2014}, and evolution of faults during bug fixing~\cite{Meulen-FaultsFailureBehaviour-ISSE-2004}.

Despite the growing interest in field faults and the design of approaches to address different kinds of faults,  
there is still no study on the nature of field problems that indicates whether they can be better addressed in-house by improving testing techniques and methodologies, or in the field by exploiting the many instances of a same application running within several heterogeneous environments. 

In this paper we present a study which provides an initial characterization of field faults and the consequent failures in the field: %
(i) we introduce a set of characteristics that make faults hardly detectable in-house, (ii) we study the characteristics of failures reported by the users from three ecosystems, and (iii) we discuss the factors that make these failures likely observable only in the field.
Our results indicate that:
\begin{itemize} [leftmargin=*]
\item 70\% of the problems observed in the field are extremely hard if not impossible to detect with in-house testing approaches, and are potentially easy to detect in the field; 
\item 78\% of the problems that are hard to detect in house can be observed only in the presence of  resources available in the field, for example new plugins, files and network connections, further emphasizing the role of the field to reveal these problems.
\end{itemize}

These results corroborate the intuition that we need more \emph{in-field software verification approaches} that exploit the resources available in the field to complement classic in-house V\&V strategies.

The analysis is based on bug reports from three ecosystems - Eclipse, OpenOffice and Nuxeo - and gives initial evidence of the predominance of field faults that can be hardly revealed in-house. The results of our analysis are publicly available at \url{http://www.lta.disco.unimib.it/tools/field/}.

The paper is organized as follows. Section~\ref{sec:fieldfailures} proposes a taxonomy of field failures. Sections~\ref{sec:rqs} and~\ref{sec:subjects} present the research questions that we investigated and the Ecosystems considered in our study, respectively. Section~\ref{sec:procedure} discusses the empirical procedure we followed to investigate the research questions. Section~\ref{sec:results} presents the results of our study about the nature and diffusion of field failures.
Section~\ref{sec:discussion} discusses the main findings,
Section~\ref{sec:related} presents related work, and
Section~\ref{sec:conclusion} summarizes the results presented in the paper. 


%% file: categories.tex
\section{Field Failures}
\label{sec:fieldfailures}

In this section, we introduce the key concepts that are relevant in our study: \emph{field failure}, \emph{field fault} and \emph{field-intrinsic fault}.

\begin{pdefinition}{Field Failure}
A \emph{field failure} is a software failure experienced in a production environment. 
\end{pdefinition}


\begin{pdefinition}{Field Fault}
 A \emph{field fault} is a fault that is present in a software program deployed and running in a production environment. 
\end{pdefinition}

Field faults may or may not cause field failures, depending on the execution conditions in the field. 

\begin{pdefinition}{In-house Faults and Failures}
The term \emph{in-house} refers to the development environment. 
\emph{In-house failures} indicate failures that occur when testing the software system in the development environment and \emph{in-house faults} indicate the causes of the failures exposed during testing. 
\end{pdefinition}

The distinction between \emph{field} and \emph{in-house} failures depends only on the time the failures are exposed and not on the nature of the fault.  The same failures may be \emph{field failures} if occurring in the field and \emph{in-house failures} if revealed during testing.  We differentiate faults by their nature by introducing the new concept of field-intrinsic faults and the corresponding taxonomy.
%
%

Field faults might be either faults that simply escape the testing phase as a consequence of an inaccurate testing process, or problems that are hard or sometime even impossible to reveal in-house before the software is executed in the field. 
Distinguishing between these two classes of field-faults is extremely important, because they call for different methods and techniques to be effectively addressed. Faults that simply escape the testing phase as a consequence of an inaccurate testing process can be addressed by improving the in-house testing and analysis activities, while faults that are hard or sometime even impossible to reveal in-house should be addressed with methods that operate in the field. 

The empirical data reported in this paper indicate three main categories of factors that harden detecting software faults in-house: faults impossible to activate in-house, faults that depend on unknown conditions, and faults that depend on "uncountable" many configurations. \emph{Faults impossible to activate in-house} are faults that depend on conditions that cannot be simulated in laboratory. 
\emph{Faults that depend on unknown conditions} are faults that are activated by undocumented situations which cannot be thus identified with any systematic approach.
\emph{Faults that depend on "uncountable" many inputs and conditions} are faults characterized by an input and configuration spaces so large that cannot be effectively addressed in-house either exhaustively or selectively. For example the huge heterogeneity of unique devices, operating systems, apps versions and configurations that characterize the mobile phone market cannot be tested exhaustively and do not present similarities that support any effective selective approach, that is, faults may be revealed in-house by chance, but would escape any feasible testing campaign no matter how effectively designed.


\begin{figure}[t!]
\centering
\includegraphics[width=7cm]{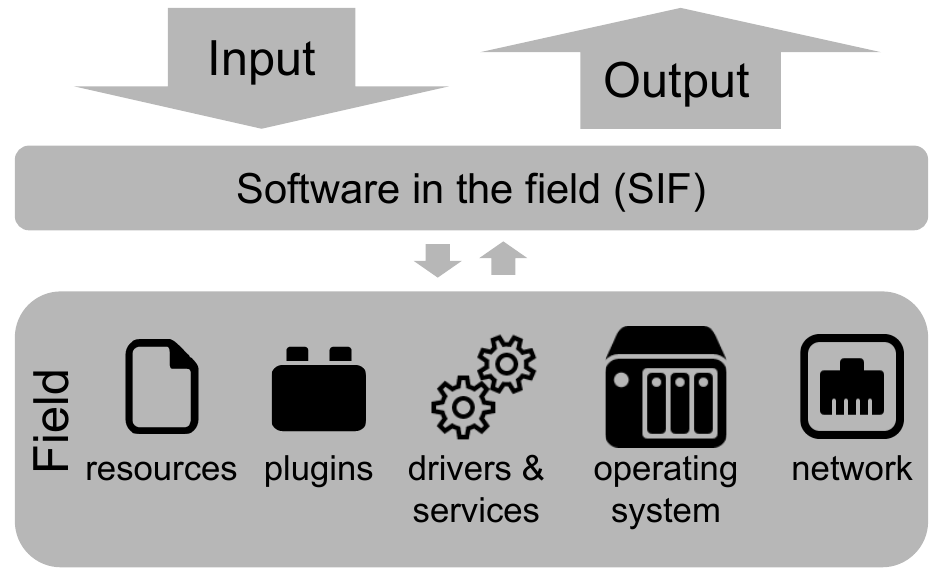}
\caption{General production environment}
\label{fig:arch}
\end{figure}

Distinguishing between faults that survive the testing due to inadequate quality processes and faults that are inherently hard to detect no matter if because impossible to activate or depending on unknown or uncountable execution conditions is important to devise verification and validation strategies. Witnessing the presence and quantity of the different classes of faults is important to define suitable V\&V campaigns. 
We start the study and capture the different nature of faults with the new concept of field-intrinsic faults that we experimentally investigate in detail in the rest of this paper.

\begin{pdefinition}{Field-intrinsic Fault}
A \emph{field-intrinsic fault} is a field fault that is inherently hard to detect in-house, either because impossible to activate in-house or because depending on unknown or "uncountable" many conditions.
\end{pdefinition}

Although some field-intrinsic faults can be revealed and removed in-house by chance, most field-intrinsic faults escape any reasonable pre-deployment V\&V activity and manifest only in the field.  In the reminder of this paper we characterize and classify field-intrinsic faults and support the need of new field V\&V activities.
We report the results of an empirical study of a body of field faults that we found in the fault repositories of different applications aiming to analyze the distribution of field-intrinsic faults in production environments. 

Figure~\ref{fig:arch} shows the different elements that comprise a production environment and that play a key role in field-intrinsic faults. 
The \emph{software in the field} (\emph{SIF}) represents a software application or a software system running in the field. The SIF receives inputs and produces outputs. The SIF receives the \emph{inputs} in the form of data and stimuli from both the SIF users and other systems interacting with the SIF. 
The SIF \emph{outputs} 
might be either visualized for the users or dispatched to other systems. While executing, the SIF may interact with a \emph{field} that includes several entities \emph{that are not under the control of the SIF}: multiple types of \emph{resources}, such as files and databases, which might be accessed by the SIF during computations, and third-party components that provide services to the SIF, such as  \emph{plugins} that extend the capabilities of the SIF with additional features, \emph{drivers \& services} that provide a range of services to the SIF, and the \emph{operating system} that defines the basic runtime environment of the SIF. 
Finally, the SIF may communicate with other applications and services using the \emph{network}. 

The role of the environment in field failures leads to the concept of \emph{failure context}:

\begin{pdefinition}{Failure Context}
A \emph{failure context} is the execution context of a failure, that is, the specific state that the elements in the field must have to trigger the failure.
\end{pdefinition}



 

%% file: researchquestions.tex
\section{Research Questions}
\label{sec:rqs}

We studied the nature and distribution of field failures by analyzing a set of bug reports produced by the end-users of three ecosystems, and
we articulated the roadmap of our study in terms of four research questions. 

\begin{researchquestion}
\RQcomplexityN: \RQcomplexity{}
\end{researchquestion}

We analyzed the field failures in the subject studies to identify the main factors that determine the faults to survive after testing and continue to exist in the field.

\begin{researchquestion}
\RQcontextN: \RQcontext{}
\end{researchquestion}

We analyzed the dependencies of field failures on the field itself, to identify which elements of the field are involved in the failures.

\begin{researchquestion}
\RQmanifestN: \RQmanifest{}
\end{researchquestion}

We clustered the field failures reported in the subject studies into classes, and identified some relevant types according to their impact and detectability.
%

\begin{researchquestion}
\RQstepsN: \RQsteps{}
\end{researchquestion}

We estimated the least subset of steps required to reproduce the field failures. 
We identify steps as user actions, being the subject studies reactive GUI applications.
 

%% file: subjects.tex
\section{Study Subjects} \label{sec:subjects}

We selected a set of desktop and web applications that (i) are available with the source code, (ii) are widely adopted and are thus good representatives of well used applications, and (iii) give access to publicly available bug reports, which are needed to study bug reports submitted by end-users.
We thus selected multiple applications from three ecosystems:

\begin{description} [leftmargin=!] 
\item[Eclipse] and in particular its well-known and widely used plugins: the \emph{Subversive} SVN client for Eclipse~\cite{Subversive}, the \emph{EGit} Eclipse Team provider for Git~\cite{EGit} and the \emph{EclipseLink} plugin for developing code using the Java persistence API~\cite{EclipseLink}. 
The bug reports are accessible on the Eclipse Bugzilla bug tracking system~\cite{EclipseBugzilla}. 
\item[OpenOffice] is one of the most popular open source office applications~\cite{OpenOffice}. The bug reports are accessible on the Apache OpenOffice Bugzilla bug tracking system~\cite{OpenOfficeBugzilla}. 
\item[Nuxeo]  is a Web-based content management system used to develop many popular Web sites~\cite{nuxeo}. The Nuxeo issue tracking is Jira~\cite{jira}. \end{description}

%% file: procedure.tex
\section{Experimental Procedure} \label{sec:procedure}

For our analysis, we identified as faults the bugs labeled as \textit{confirmed}, \textit{verified} or \textit{resolved}, and we inspected all the bugs reported for the three Eclipse plugins from January 2015 to December 2015 for a total of 412 analyzed bug reports, and all the bugs reported for both OpenOffice and Nuxeo from September \nth{1} 2016 to October \nth{1} 2016, for a total of 99 and 56 bug reports inspected, respectively. 

For each bug report, we inspected the information about the failure, the inputs, the execution conditions and the failure impact. 
We discarded the bug reports containing only a memory dump and a stack trace, which might be useful for developers, but are not useful for the purpose of our investigation, and studied in detail a total of 119 bug reports: 63 for Eclipse, 26 for OpenOffice, and 30 for Nuxeo. 




\smallskip
\noindent \textbf{\RQcomplexityN: \RQcomplexity{}}

We investigated why faults have not been revealed in house but have been detected only in the field by examining the conditions that caused the failures to identify the factors that contribute to the failures and are extremely hard to be tested in-house. 
We labeled each fault as \emph{bad-testing}, if we could not find any of such factors,  \emph{field-intrinsic} otherwise. 
We identified four categories of \emph{field-intrinsic} faults that
we discuss in the next section, where we characterize the identified classes of faults, and we report both qualitative and quantitative data. 
We used only the faults labeled as \emph{field-intrinsic} to answer the other three research questions.



\smallskip
\noindent \textbf{\RQcontextN: \RQcontext{}}



For each bug report, we identified the elements of the field shown in Figure~\ref{fig:arch} that play an essential role in the failure. 

\smallskip
\noindent \textbf{\RQmanifestN: \RQmanifest}
\input{tableFailureTypes}

We studied the characteristics of field failures to identify their attributes and classify them.
Better understanding the nature of field failures is essential for developing techniques for testing applications in the field without uncontrolled side effects. 
Some types of failures might be easier to detect and control than others. 
For example, exception and error messages are easy to detect and usually do not cause loss of data because the application itself detects and handles these erroneous situations; system crashes are also easy to detect, but may cause loss of user data; incorrect results may be hard to detect, and may silently compromise the user data and the overall computation. 

We carefully analyzed the failure taxonomies proposed by Bondavali and Simoncini~\cite{Bondavalli-FailureClassification-FTDCS-1990}, Aysan et al.~\cite{Aysan-ErrorModeling-COMPSAC-2008}, Avizienis et al.~\cite{Avizienis-FailureTaxonomy-TDSC-2004}, Chillarege et al.~\cite{ChillaregeTSE1992}, and Cinque et al.~\cite{Cinque-MobilePhoneFailureTaxonomy-DSN-2007} to identify the candidate attributes for field failures, and exhaustively inspected the bug reports in our data set to identify the most relevant attributes for characterising field failures: \emph{failure type} and \emph{detectability}. 




\smallskip
\emph{\textbf{Failure Type}}
The failure type characterizes a failure according to the way it appears to an observer external to the system. 
\smallskip

We identified three possible categories of failure types, \emph{value}, \emph{timing} and \emph{system} failures, and we further detailed each type in three subtypes, for a total of nine failure types, which we use in the next sections to categorize bug reports, and which are summarised in Table~\ref{table:failuretypes}.

\begin{description} [leftmargin=!] 
\item[Value failures] occur when the SIF produces incorrect outputs: an \emph{invalid value}, a \emph{value out of domain} or an \emph{error message}.
For example in a functionality that returns the ZIP code of a city, a \emph{value failure} of type \emph{invalid value} occurs when the SIF returns the ZIP code associated with a city different than the input one, a \emph{value failure} of type \emph{out of domain} occurs when the SIF returns a malformed ZIP code, a \emph{value failure} of type \emph{error message} occurs when the SIF returns a message the reports an internal error that prevented retrieving a ZIP code.

\item[Timing failures] occur when the SIF produces some outputs at a wrong time: too early (\emph{early timing}), too late (\emph{late timing}) or never (\emph{omission}).

\item[System failures] occur when the SIF is blocked (\emph{halting failure}) has stopped running (\emph{crash}) or does not respond reliably to the input stimuli (\emph{unstable behavior}).
\end{description}

\smallskip
\emph{\textbf{Detectability}}
The detectability attribute characterizes the  difficulty of detecting the failure. 

\input{tableDetectability}

We distinguish four levels of detectability, \emph{signaled}, \emph{unhandled}, \emph{silent} and \emph{self-healed}, based on both the ability of the system to detect the failure and an external observer to observe a misbehavior without specific system knowledge, as summarized in Table~\ref{table:failuredetectability}.

\begin{description}[leftmargin=!] 
\item [Signaled failure:] a failure that the system detects and reports. 
A simple example of a signaled failure is an application that opens a popup window to inform the user that the application will be unexpectedly closed because of a memory problem; 
\item[Unhandled failure:] a failure that the system does not handle and that leads to a crash. The system does not detect the failure, while the user trivially detects the uncontrolled crash of the application without requiring any knowledge about the application; 
\item[Silent failure:] a failure that the system does not detect letting the application continue operating 
without producing any signal that a user can recognize as a failure without prior knowledge about the application. 
A simple example of silent failure is a flight simulator that simulates the flight conditions imprecisely and that a user cannot detect without a specific knowledge of the flight simulation system.
\item[Self-healed failure:] a failure that the system detects and overcomes transparently to the user. The user continues using the application without noticing any problem. Self-healed failures are common in systems exploiting redundancy to mask failures, such as Hadoop~\cite{Hadoop}.
\end{description}

\smallskip
\textbf{\RQstepsN: \RQsteps{}}

For each failure, we identified a sequence of steps that are needed to cause the failure, aiming to, but not necessarily proved to be, a minimal sequence.  
For the interactive subjects, we identify steps with GUI actions like opening windows, entering data in some fields, clicking on menus and buttons.

We counted the steps that lead to a failure by considering the sequence of operations 
\CHANGED{described}
in the bug reports submitted by users. 
When creating a bug report, users intuitively identify a critical state that may lead to the failure and submit both the information about the critical status, typically described in a declarative way, and a sequence of operations that lead to a system failure from the critical state, typically described in an operational way. 
For example in the Open Office bug report \#126930, the state to trigger the failure is characterized by the availability of a certain file, and the steps to reproduce the failure consist of opening the file, scrolling the document, selecting a frame, and enlarging the frame. 
We identified the minimal subset of the actions reported by the user that are needed to cause the failure from the critical state indicated in the bug report, which often corresponds to the minimal number of actions needed to reproduce the failure~\cite{roehm2015automated}. 

The amount of steps needed to reproduce a failure is an important information for estimating the complexity of testing techniques that work in the field and reveal failures by monitoring the status of the application to detect failure-prone states and executing test cases of appropriate complexity  when a failure prone state is detected.  

%% file: tableFailureTypes.tex
 \begin{table}
   \centering
   \caption{Failure Types} \label{table:failuretypes}

\begin{footnotesize}
{\setlength\extrarowheight{2pt}
\begin{tabular}{ p{2.1cm}  p{5.8cm} }

\multicolumn{2}{c}{\emph{\underline{Value Failures}}}\\
Invalid value & The SIF produces an incorrect output value, although still in the domain of the output variable\\ \cdashline{1-2}
 Out of domain & The SIF produces a value outside the domain of the output variable\\ \cdashline{1-2}
  Error message & The SIF produces an error message\\

\hline

\multicolumn{2}{c}{\emph{\underline{Timing Failures}}}\\

Early timing & The SIF produces an output too early, for instance before an expected waiting time\\  \cdashline{1-2}%
  Late timing & The SIF produces a value after a required deadline, defined either explicitly or implicitly\\ \cdashline{1-2}
  Omission & The SIF never produces an output value in an asynchronous computation\\

\hline

\multicolumn{2}{c}{\emph{\underline{System Failures}}}\\

Halting failure & The SIF never produces any output value in a synchronous computation\\ \cdashline{1-2}
  Crash & The SIF crashes and no services is delivered\\ \cdashline{1-2}
  Unstable behavior & The SIF shows an erratic behavior without receiving any input, for instance, a flashing blacklight in a smartphone\\

\end{tabular}
}
\end{footnotesize}

\vspace{-0.5cm}
\end{table}

%% file: tableDetectability.tex
 \begin{table}
   \centering
   \caption{Failure Detectability} \label{table:failuredetectability}
\begin{tabular}{ll}
 \begin{minipage}[b]{0.32\hsize}\centering
\begin{tabular}{l  | c | c }
 &User  &  SIF \\ \hline

Signaled & $\surd$ & $\surd$  \\ \hline

Unhandled & $\surd$ & $\times$\\ \hline

Silent & $\times$ & $\times$ \\ \hline
 
Self-healed & $\times$ & $\surd$ \\

\end{tabular} 
\end{minipage}
&
\begin{minipage}[b]{0.32\hsize}\centering
\begin{footnotesize}
LEGEND::\\
\begin{tabular}{cp{7.5cm}}
$\surd$ & detected by \\
$\times$ & NOT detected by\\   
\end{tabular}
\end{footnotesize}

\end{minipage}
\end{tabular}
\vspace{-0.5cm}
\end{table}

%% file: results.tex
\section{Results}
\label{sec:results}


\subsection*{\RQcomplexityN: \RQcomplexity{}}
%

We analyzed the bug reports to distinguish faults that are due to insufficient testing (\emph{bad testing} (BT)) from \emph{field-intrinsic faults}.
We further analyzed the field-intrinsic faults and identified four types of conditions that lead to field-intrinsic faults and that we use to classify such faults: \emph{Irreproducible Execution Condition} (IEC), \emph{Unknown Application Condition} (UAC), \emph{Unknown Environment Condition} (UEC), and \emph{Combinatorial Explosion} (CE).
%
%
%
The identified classes of faults 
comprise a complete taxonomy for the faults in the bug reports that we analyzed, and represent an initial general framework for classifying field faults.
%
 
\subsubsection*{Irreproducible Execution Condition (IEC) Faults} 
IEC faults are faults that can be revealed only under conditions that cannot be created in-house. This may depend on the impossibility of reproducing the complexity of the whole field environment, the inability of creating the specific failing execution or the evolution of the environment and the interactions with the SIF.

The safety critical routines to be executed in the case of natural disasters are good examples of execution conditions that might be impossible to reproduce in-house. Although a disaster can be simulated to some extent, a major natural disaster, for instance an earthquake or a tsunami, cannot be fully reproduced in-house, and some field-intrinsic faults may depend on extraordinary combinations of events that can be observed only in real conditions. 

Similarly, the behavior of a system for an increasing number of users who interact with the application according to patterns that are not entirely predictable is often hard to test, especially for extreme situations, such as the extraordinary online streaming services workload experienced in the Super Bowl night~\cite{superbowlAppNotworking}.

The evolving varieties of configurations, for instance versions of the operating systems, drivers and plugins, are good examples of unpredictable changes to interactions between the SIF and the environment (hereafter \emph{SIF-environment interactions}).
New versions or entirely new plugins or drivers distributed after the most recent SIF release might generate faults impossible to reveal in-house before the release itself. 
%
%
%
%

An example of such situation is the fault described in the EclipseLink bug report \#429992. 
EclipseLink is an Eclipse plugin for developing Java applications that uses JPA to map objects into databases. The bug report indicates that EclipseLink silently ignores classes that contain lambda expressions: even if an object should be persisted in the database because its class includes the \emph{@Entity} annotation, no table for persisting the object is generated in the database. 
Since lambda expressions have been introduced only in Java 8, it was impossible to test the combination of lambda expressions with JPA annotations when the EclipseLink plugin was developed, before the release of Java 8. 
EclipseLink should not have been affected by the presence of lambda expressions and should have supported the persistency of the classes regardless of the presence of lambda expressions. 
However due to an unforeseen compatibility issue, EclipseLink stopped working correctly when processing classes with lambda expressions.


\subsubsection*{Unknown Application Condition (UAC) Faults}
UAC faults are faults that can be revealed only with input sequences that although executable in-house depend on conditions about the application that are ignored before the field execution and thus cannot be captured in in-house test suites.  

An example of field failures that derive from unknown conditions is the Eclipse Subversive report \#459010, which indicates that Subversive fails when retrieving folders whose name terminates with a blank character. This corner case is not documented in the specifications, and is hard to reveal with in-house testing because of the lack of information that may suggest to design test cases covering this specific situation. Structural test suites do not address this problem either, since many problems of this type are due to missed code, as in the case of this fault.

Another example of a UAC fault is the Eclipse \#440413 bug report, which describes a fault in method \texttt{convertObjectToString} of class \texttt{XMLConversionManager} that converts any object to a proper string representation. 
The method works properly except when used to convert a \texttt{Big\-Decimal} representing a number in scientific notation, since it returns a string that encodes a number in scientific notation and not a plain number as expected. 
We verified that this case is not mentioned either in the \texttt{XMLConversionManager} specification or in the API documentation, and is thus hidden to the testers who did not reveal the bug during testing and discovered it in the field after the software has been released.

In our experimental analysis, we did not have always access to the specification of the software. When this happened, we classified a fault as UAC when the inputs that lead to the failure are largely unrelated with the purpose of the functionality that fails, assuming that such cases were not defined in the specifications.
Thus our classification may not be perfectly accurate.

\subsubsection*{Unknown Environment Condition (UEC) Faults}
UEC faults are faults that can be revealed only with information about the environment that is not available before field execution. 
UEC faults are hardly detectable with in-house test cases designed without a complete description of the constraints on the SIF-environment interactions.

The full range of behaviors of third-party services that the SIF accesses through the network is a good example of information rarely completely available at design time, and thus possible source of UEC faults.
 An example of UEC fault is the Eclipse bug report \#394400 that indicates that EclipseLink may fail with a \texttt{NullPointerException} when executed under heavy load on the Oracle JRockit VM.
The issue depends on the behavior of the Just In Time compilation feature of the JRockit VM that may reorder the operations executed within method \texttt{isOverriddenEvent} so that it returns  an incomplete result. 
This undocumented behavior is responsible for the EclipseLink exception. 



\subsubsection*{Combinatorial Explosion (CE) Faults}
Even when the behavior of both the application and the environment are fully specified and can be replicated in-house, the combination of the cases to be tested may increase to a magnitude of cases that cannot be fully tested in-house. 
There are many sources of combinatorial explosion in software applications, such as the many possible configurations and preferences, the combinations of inputs and states, the many environments, for instance operative systems and hardware devices, that can be used to run an application, and so on. 
A well known example of combinatorial explosion of combinations are the sets of hardware devices, operating systems and configurations that comprise the execution conditions of smartphone applications that can almost never be fully tested in-house.


%
%
%
%

An example of a CE fault is the fault described in the Eclipse bug report \#484494, which indicates that the diff feature of the Subversive plugin does not work when comparing a file to a symlink of a file that has been moved. 
Changing the location of a file referred by a symlink and using the symlink as part of a comparison is a legal combination of operations among the huge set of combinations that comprise to the sequence: $\langle$change the status of a resource, use the changed resource as part of a computation$\rangle$. 
Systematically testing all these combinations commonly exceed any reasonable albeit impressively large testing budget, because of the many ways resources can be changed independently from each other. 


In our analysis, we observed that only a small percentage of CE cases are due to specific inputs (18\% of the cases), while the rest of the CE cases are due to field elements.

\subsubsection*{Bad Testing (BT) Faults}

We conclude our taxonomy with a discussion of BT faults, which we classify in our experimental analysis as \emph{field} but not \emph{field-intrinsic} faults.
BT faults are faults that are not detected in-house due to weaknesses of the testing process. 
We include in this class all the faults in the field that do not belong to any of the previously described classes. 






An example of BT fault is the fault reported in the Subversive bug report \#326694, which indicates that Subversive erroneously reports as conflicting two identical files that have accumulated the same set of changes on two different branches. 
Since detecting conflicts is a primary feature of this 
plugin, developers should have tested a basic case like the presence of the same changes in two distinct branches. 

\subsubsection*{Taxonomy}

The taxonomy that we proposed in this paper opens new scenarios of increasing complexity.  
BT faults simply substantiate the need of improving the in-house testing process and do not introduce new challenges for the software testing community. UAC, UEC and CE faults call for new techniques to enrich well designed test suites with test cases identified in the field while experiencing faults caused by unpredictable (UAC and UEC) or impossible-to-exhaustively-test (CE) conditions. The main challenge that has been only partially addressed so far is to record execution sequences that lead to failures in the field, and reproduce them either in the field or in house to identify and remove the faults. 
Being not executable in house, IEC faults further challenge the software testing community with the problem of executing failing test cases in the field.
The main challenges are to both reveal failures by executing test cases in the field, which requires to control the execution of the test cases in the usually complex field context, and prevent any side-effect for the users. 

\subsubsection*{Quantitative analysis}

\begin{figure}
\centering
\includegraphics[width=8.4cm]{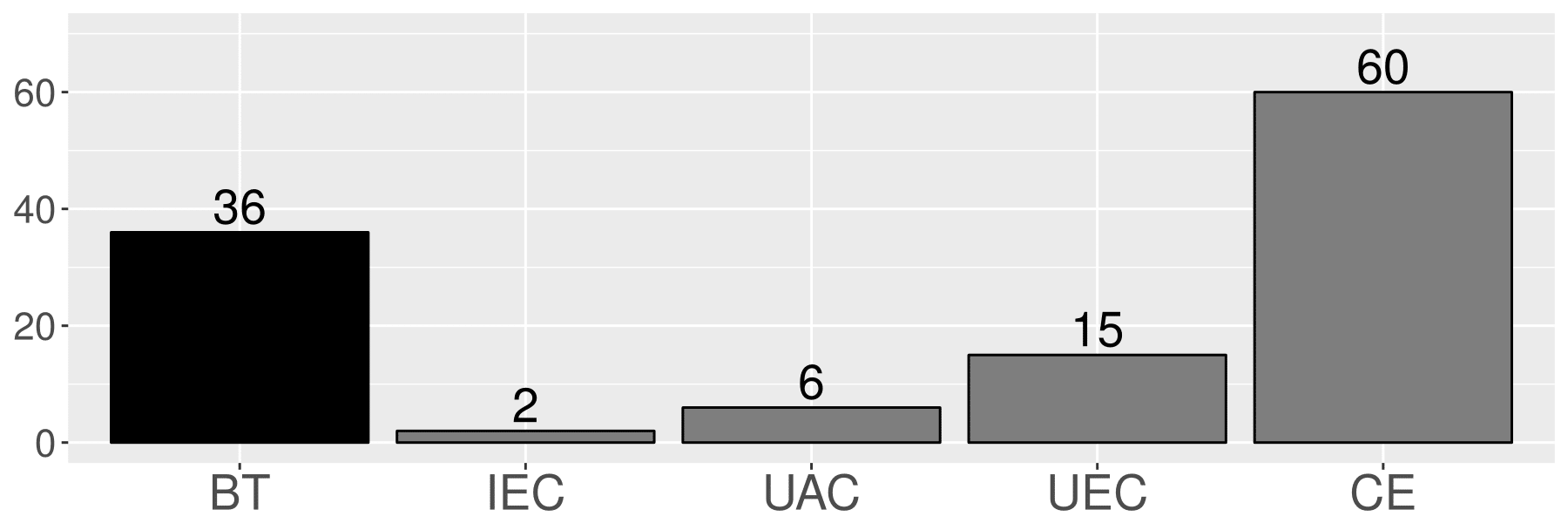}
\caption{\RQcomplexityN: \RQcomplexity{}}
\label{fig:rq1}
\end{figure}

Figure~\ref{fig:rq1} summarizes the quantitative results of our empirical investigation. 
The bar chart indicates the number of faults classified in the five categories discussed above, and shows that field-intrinsic faults (the sum of the IEC, UAC, UEC and CE columns) are the majority of the field faults in our data set.
Field-intrinsic faults represent 70\% of the analyzed bug reports, thus confirming that field faults cannot be addressed by simply enhancing the testing process, but calls for specific in-field approaches.


The bar chart indicates that \emph{combinatorial explosion} (CE) is the most frequent cause of field-intrinsic faults, while \emph{Irreproducible Execution Condition} (IEC) is the least common source of faults. 
Unknown execution conditions of either the application or the environment (UAC and UEC faults) are also relatively frequent cases. 
The dominance of CE faults is not surprising: The behavior of SIFs is influenced by many factors that can be never exhaustively tested in house. 
The many combinations that are hard to design, foresee and test in house, can be order of magnitudes easier to address in the field, where such a diversity is spontaneously and implicitly available. 

Our analysis identified few \emph{Irreproducible Execution Condition} (IEC) faults, all caused by evolution of the SIF-environment interactions that emerged after the deployment of plugins not available at the time of testing, before the deployment of the SIF in the field. 
The scarce presence of IEC faults may depend on the nature of the applications that we analyzed. 
In other domains the presence of IEC faults might be higher. 
Consider for instance the domain of embedded software, where the interactions with the physical world might be sometime extremely hard to test. 
\begin{changed}
We observed a similar trend for the three subjects: a predominance of CE faults (openoffice being the highest at 73\%) and a total of 10 - 20 \% faults falling into the UEC/UAC category. 
It is worth noting that the two IEC faults we identified were both on the Eclipse platform.
\end{changed}




\subsection*{\RQcontextN: \RQcontext{}}

We analyzed the bug reports to study the role of field elements in field failures, and validate the intuitive hypothesis that many field-intrinsic faults may be hard to reveal in-house because their activation may depend on one or more elements that should be present in the field and should be in the right status to produce the failure. 
Below, we discuss the role played by the field elements that we introduced in Figure~\ref{fig:arch}, provide \CHANGED{concrete} examples, 
and discuss the quantitative data from the experimental data sets.
%

\paragraph{\textbf{Resources}}

Software applications typically interact with many resources during the computation. 
For instance, many applications read from and write to persistent units, such as files and databases. 
Causes of field failures may involve resources in many ways. In our investigation, we observed two main cases: interactions between SIF and resources (hereafter \emph{SIF-resource interactions}) that lead to performance problems and SIF-resource interactions leading to functional problems. The unbearable amount of time for SIFs to process some large resources and SIFs incorrectly handling resources of a particular type are examples of performance and functional problems triggered by SIF-resource interactions, respectively.
 
An example of SIF-resource interaction that triggers a performance problem is described in the OpenOffice bug report \#95974. 
The OpenOffice writer crashes when trying to open a \texttt{.odt} document longer than 375 pages. The failure causes the CPU usage and the disk access rate to increase to $100\%$, and the application window simply crashes after one minute of unresponsiveness, activating the recovery wizard.





\paragraph{\textbf{Plugins}}

The plugin mechanism is a common solution to extend applications with new functionality in the field. 
In the presence of plugins, applications work as operating systems that embed the plugin executions, and interact with the plugins to access specific functionalities.
Applications and plugins are developed and maintained independently. Evolution at either sides may trigger failures due to unforeseen interactions. 

%
%
%

For example, the EGit bug report \#383376 indicates that the repository search does not work on Github due to an unforeseen interaction with the Mylin Github connector plugin. 


\paragraph{\textbf{Operating system}}

Many applications can be executed on different versions of different operating systems. The interactions of a SIF with a specific version of an operating system may trigger failures otherwise unexperienced.  

%
%

An example of a problem involving the operating system is the failure documented in the OpenOffice bug report \#126622 that describes how the OpenOffice writer does not correctly handle functionalities involving tables and queries under OSX. The failure prevents OpenOffice from closing, and forces the users to restart the operating system.

\paragraph{\textbf{Drivers and services}}

Applications often interact with third party drivers and services, whose availability depend on the production environment. During in-house development specific combinations might remain untested and failures unrevealed.




For example, the fault documented in the Eclipse Egit bug report \#435866 indicates that the Eclipse Egit version control system fails to open the required network connections due to some unexpected changes of the authentication methods implemented in the Eclipse connection service.

\paragraph{\textbf{Network}}

Many software applications use the network to access resources or functionalities that are not available locally. With a plethora of different network protocols available, failures might be triggered when an application uses a specific protocol.

%
%
%


For instance, the fault described in the Nuxeo bug report \#20481 describes a failure caused by a connection timeout that occurs when users download big zip files. Nuxeo does not handle connection timeouts properly and does not clean up temporary files, which leads to resources exhaustion.


\CHANGED{\paragraph{\textbf{None}} In a few cases the field-intrinsic faults do not depend on any interaction between the field elements and the SIF. Although not depending on any field element, these faults are still extremely hard to reveal at testing time, for instance because they can be revealed only by selecting a specific input out of a combinatorial number of cases.}

This is the case of the OpenOffice bug report \#126953, which indicates that when changing the format of a paragraph wrIECen with the Verdana font to italics bold, OpenOffice incorrectly adds blank lines before each occurrence of the brackets '(' and ')', and the text within the brackets disappears. This failure can be triggered only with a specific combination out of millions of input combinations: the use of Verdana font, and the presence of brackets when changing fonts to italics bold, out of the many combinations of font types, characters and font properties.

\subsubsection*{Quantitative analysis}
\begin{figure}
\centering
\includegraphics[width=8.4cm]{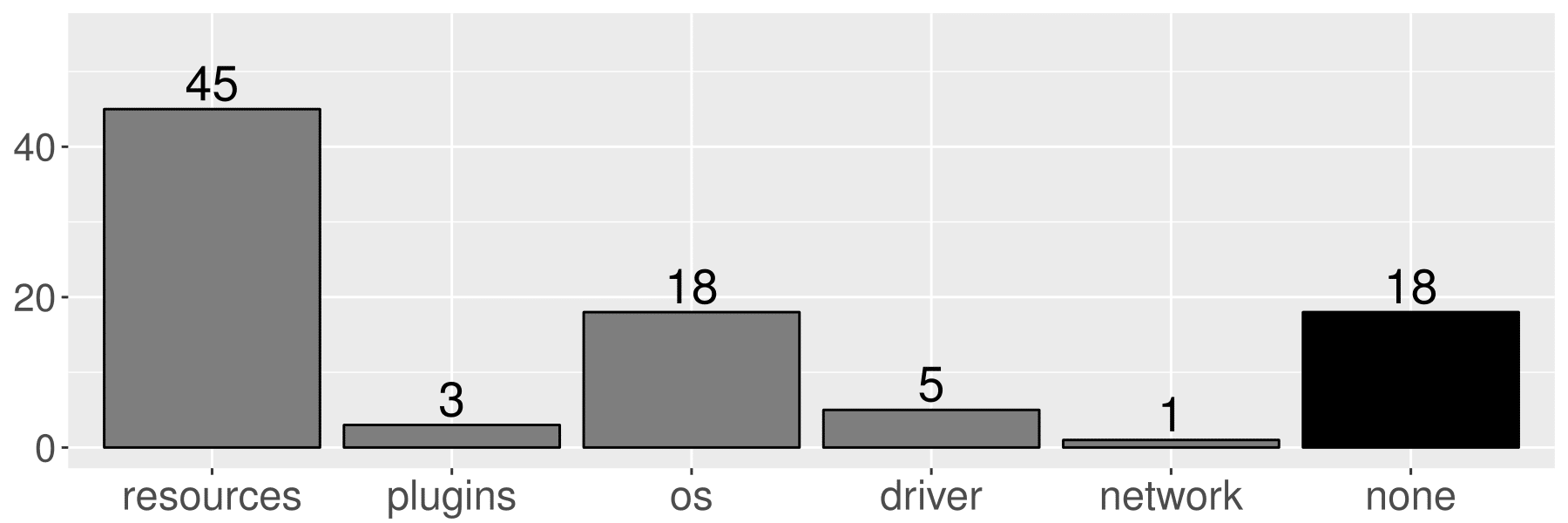}
\caption{RQ2: field elements involved in the failure}
\label{fig:rq2}
\end{figure} 

Figure~\ref{fig:rq2} quantifies the impact of the different field elements on faults, by indicating the amount of faults affected by each type of field element.
The causes are not exclusive, since a same fault may involve multiple field elements. 
In Figure~\ref{fig:rq2}, bar \emph{none} reports the number of bug reports that describe failures that do not involve any field element.

Our analysis shows that interactions with the resources are the main cause of field-intrinsic faults (49\% of the cases). 
%
Interactions with the operating systems are also a relevant cause of field-intrinsic faults (20\% of the cases).
Network, drivers \& services, and plugins have been all observed as causes of field-intrinsic faults at least once, but they are collectively observed in a small proportion of the cases (10\% of the cases in total). In total, 78\% of the field-intrinsic faults interact with a field element.

Although the data reported in Figure~\ref{fig:rq2} may be biased by the experimental setting, they already provide important information to define a research road map in the study of techniques to reveal and fix field-intrinsic faults.

\subsection*{\RQmanifestN: \RQmanifest}

\begin{figure}
\centering
\includegraphics[width=8.4cm]{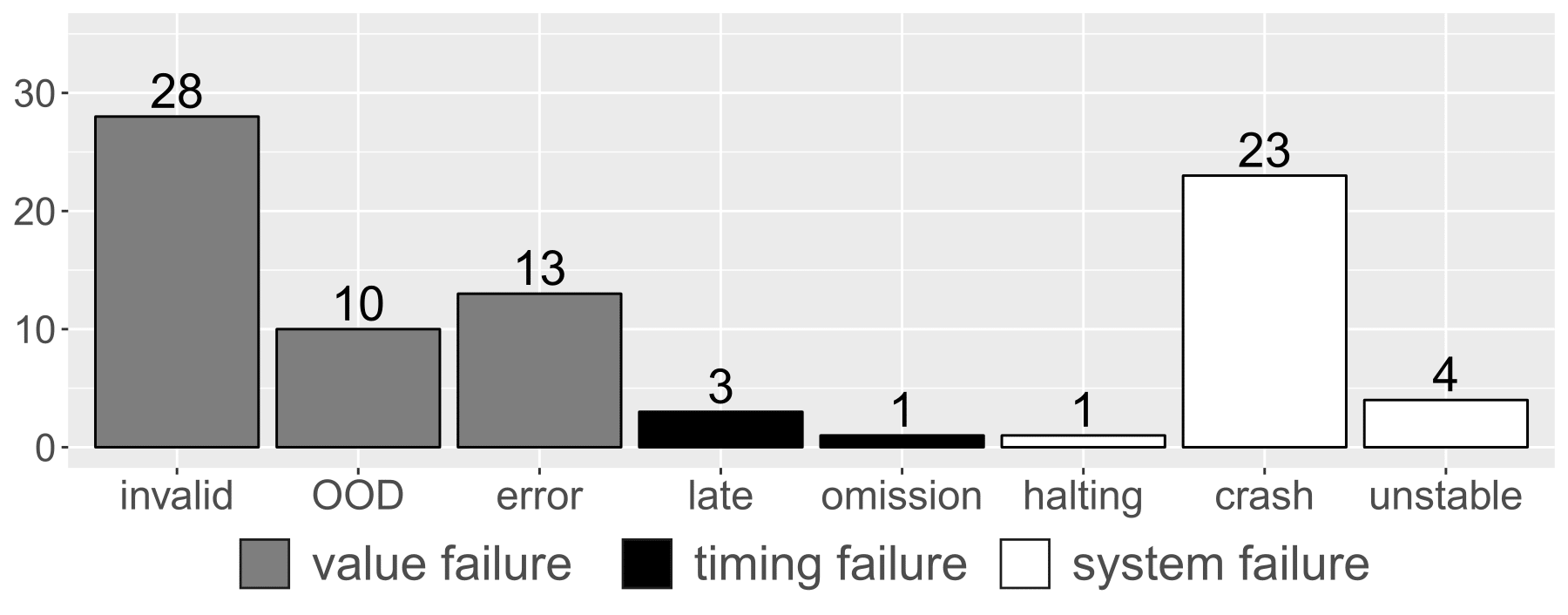}
\caption{RQ3 - Failure Type}
\label{fig:rq4}
\end{figure}

We analyzed the distribution of failure types, and investigated the issues related to detectability.
Figure~\ref{fig:rq4} plots the distribution of the failure types presented in Table~\ref{table:failuretypes}.

Most failures (51 out of 83 failures corresponding to 61\% of the analyzed failures) are \emph{value failures}, that is, executions that produce incorrect results. 
The most frequent case of  \emph{value failures} is the generation of invalid outputs, followed by the generation of error messages and the production of 
\CHANGED{values out of domain (OOD in Figure~\ref{fig:rq4})}.
System failures are also frequent (28 out of 83 failures corresponding to 34\% of the analyzed failures).
They mostly lead to system crashes, and only occasionally to either unstable behaviors or system halt.
Only a small set of the failures that we analyzed are due to the timing aspect (4 out of 83 failures corresponding to 5\% of the analyzed failures). We observed few late timing and omission failures, and no early timing failures. 

The results indicate that the generation of incorrect values (either invalid values, values out of domain or error messages) and systems crashes are the main classes of field failures (they represent 74 out of 83 failures corresponding to 89\% of the analyzed failures).
These results, and in particular the low frequency of timing failures, might depend on the domain that we investigated (desktop applications extensible with plugins and Web applications).
We expect different frequencies of failure types in other domains: In particular, we expect an increasing frequency of timing failures in embedded systems, where the synchronization among the software components plays a relevant role.
%

\smallskip

\begin{figure}
\centering
\includegraphics[width=8.4cm]{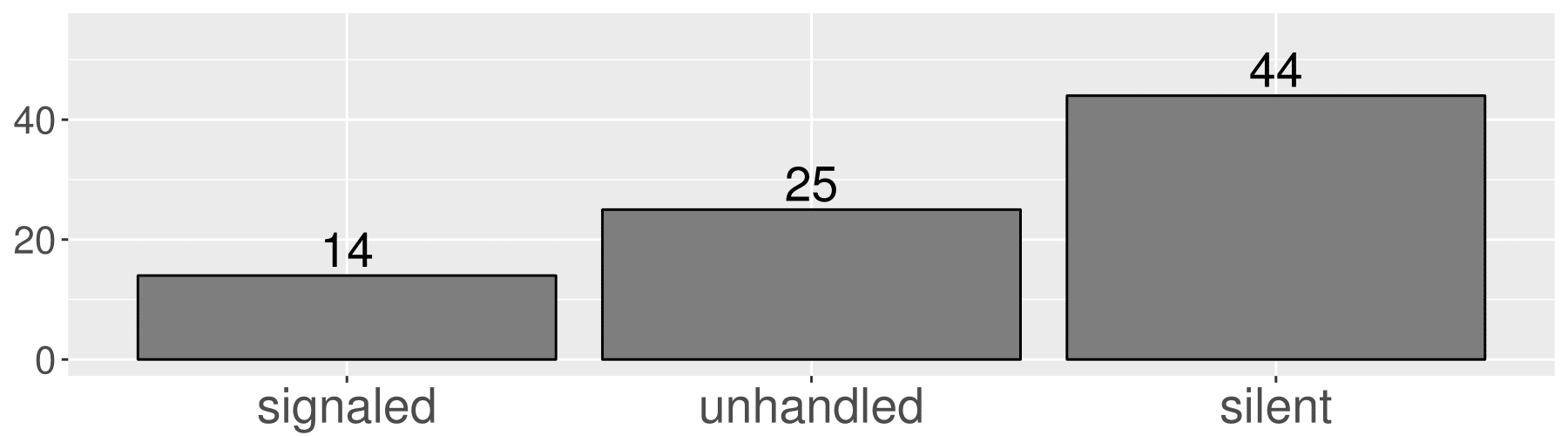}
\caption{RQ3 - Detectability}
\label{fig:rq4detect}
\end{figure}

Figure~\ref{fig:rq4detect} plots the distribution of failures by detectability according to the classes presented in Table~\ref{table:failuredetectability}. 
A relatively high portion of failures are detected because the failures are either \emph{signaled} by the application itself (14 out of 83 failures corresponding to 17\% of the analyzed failures) or \emph{unhandled} (25 out of 83 failures corresponding to 30\% of the analyzed failures) causing a system crash.
Such failures 
\CHANGED{can be easily detected.}
On the contrary, \emph{silent} failures (44 out of 83 failures corresponding to 53\% of the analyzed failures) are 
\CHANGED{hard to detect}
without some specific knowledge about the expected behavior of the application in response to certain stimuli, pointing to the well known oracle problem~\cite{Barr:OracleSurvey:TSE:2015}.
These results suggest that testing strategies working in the field without 
\CHANGED{exploiting domain specific oracles}
could hardly reveal more than half of the field-intrinsic faults.

The considered subjects do not include mechanisms to automatically overcome from failures at runtime, and thus we have not observed any occurrence of \emph{self-healed} failure.

%
%
%

\subsection*{\RQstepsN{}: \RQsteps}

\begin{figure}
\centering
\includegraphics[width=8.4cm]{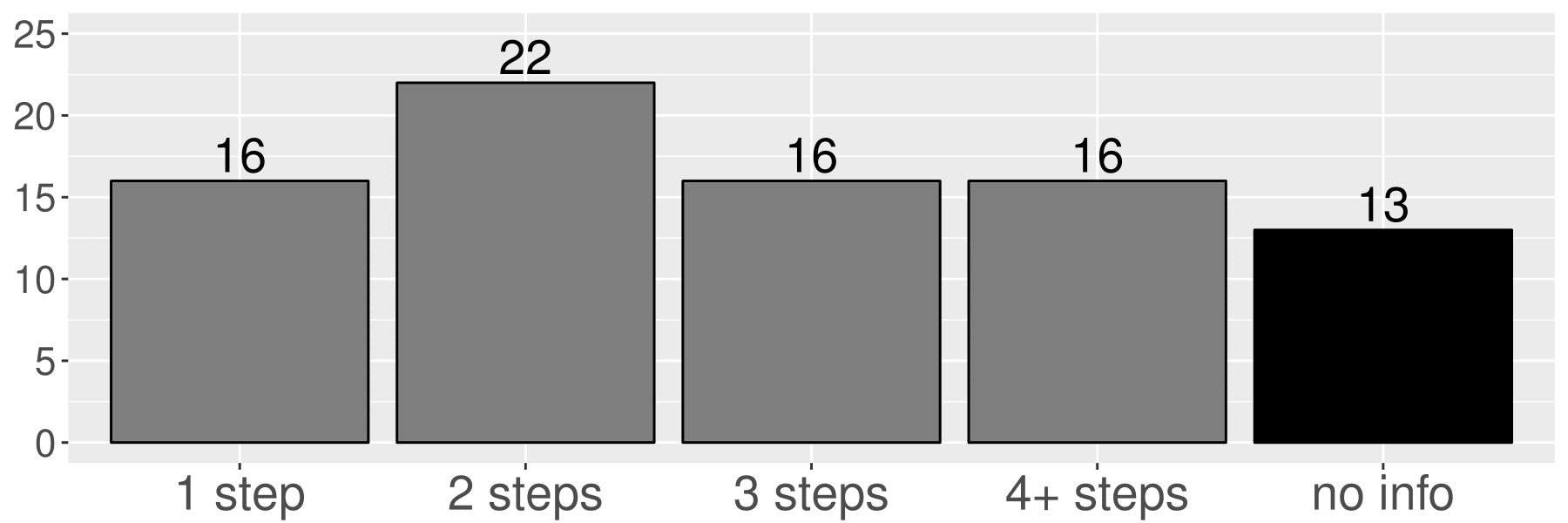}
\caption{RQ4: Steps required to trigger a failure}
\label{fig:rq3}
\end{figure}

\CHANGED{
As discussed in Section~\ref{sec:procedure}, we computed the number of user actions necessary to trigger the failures by considering the operations that are described in the bug reports limiting to the ones essential for reproducing the failure.}

Figure~\ref{fig:rq3} plots the distribution of the field-intrinsic faults by the number of steps required to reproduce the failure.
We were not able to determine the number of steps required to produce the failure in 13 out of the 83 analyzed failures (16\% cases corresponding to bar \emph{no info}), while we determined the number of steps required to reproduce the failure in 70 out of the 83 analyzed failures (84\%), and observed that a large amount of failures can be reproduced with no more than three steps (54 out of 70 reproducible failures corresponding to 77\% of the reproducible failures.) 

These results provide useful information when designing field-testing approaches, since they suggest that only few actions are necessary to reproduce a failure once reached a failure prone state, and indicate that field testing strategies should focus more on detecting failure-prone states than on generating long action sequences to reproduce the failures.

%% file: threats.tex
\subsection*{Threats to validity} \label{sec:threats}

We collected our experimental data from the bug reports of desktop applications extensible with plugins (Eclipse and OpenOffice) and Web applications (Nuxeo) by examining a limited although reasonable amount of bug reports.
\begin{changed}The results give early evidence of the nature of the failures that can be experienced in plugins and Web applications, and need further studies to be generalized to other kinds of applications and to be quantitatively assessed.\end{changed}

We defined the classification schema, and analyzed the bug reports 
manually. 
Two authors have independently analyzed the bug reports, and all the authors have discussed the conflicting cases until reaching a consensus.
Although the process we followed should mitigate the risk of misinterpretation of the cases, we cannot fully exclude clerical errors in our analysis. 
The row data and the detailed material that we refer to in the paper are publicly available for independent inspections and further uses.

The bug reports that we examined might be inaccurate some times. They may for example include partial information about the failures. 
Although we cannot fully eliminate this potential issue, we believe that possibly incomplete bug reports considered in the experiments may have reduced the number of field-intrinsic faults that we identified, thus only pessimistically affecting the results. 
In particular, the lack of information about a failure may have increased the chance of a fault to be erroneously classified as irreproducible execution condition, while the unknown conditions about the application or the environment may have reduced the amount of faults classified as combinatorial explosion faults. 
We assume that our results that indicate a density of 70\% of field-intrinsic faults among the analyzed bug reports is a conservative under approximation of the field-intrinsic faults that are present in the examined applications.


%
%
%

%% file: discussion.tex
\section{Findings} \label{sec:discussion}

The experimental data that we collected to answer the research questions lead to some interesting findings:

\finding{Most of the failures that can be observed in the field are caused by field-intrinsic faults} 
Our experimental data indicate that about 70\% of the field failures that we analyzed are caused by field-intrinsic faults, that is, are caused by faults that might be hardly revealed in house. 
These faults are caused by four challenges: combinatorial explosion, unknown environment or application condition, and situations impossible to reproduce. 
This result calls for approaches that can deal with these classes of failures in the field.

\finding{Combinatorial explosion is a relevant cause of undetected field-intrinsic faults} 
Combinatorial explosions are notably hard to address in testing and analysis techniques. 
Our experimental investigation indicates that, \begin{changed}despite numerous techniques developed to tackle the problem of generating test cases that adequately cover interactions of parameters in a software application~\cite{lei2008ipog,nie2011survey}\end{changed}, combinatorial explosion \begin{changed}still\end{changed} plays a prominent role in 
\CHANGED{preventing the detection of} 
field-intrinsic faults. 
Differently from other contexts, in the case of field-intrinsic faults, the source of combinatorial explosion is not the user input (only 18\% of the failures are caused by specific \CHANGED{combinations of} inputs) but the status of the field elements.

\finding{The interaction with the environment is almost always a relevant factor in field-intrinsic faults} 
The vast majority of the field-intrinsic faults (78\% in our study) requires some forms of interactions with the environment to be activated. 
Resources and operating systems are the most relevant field elements involved in field-failures, but also drivers, plugins and the network are often important. 
This result indicates that techniques to reveal field-intrinsic faults must take into consideration the production environment in which the system is executed.

\finding{Value and system field-faults are more frequent than timing field-faults} The ability to analyze the output produced by a system, including the ability to detect crashes, is sufficient to detect most of the field-intrinsic failures, with a rate of timing field failures as low as 5\% of the cases.

\finding{The oracle problem affects about half of the field-intrinsic faults} 
Our experimental  analysis indicates that 43\% of the failures can be detected by intercepting unhandled events, for example system crashes, and error messages. 
\CHANGED{Domain specific} oracles are necessary to address the remaining 57\% of the cases. 
\CHANGED{This calls for techniques and methods to derive strong automatic oracles for field testing.}

\finding{Field failures can be commonly revealed with short sequences of actions} 
Our experimental analysis provides evidence that few steps (three or fewer actions in 77\% of the cases) are usually needed to make the SIF fail from a failure-prone state. 
This suggests that detecting states that offer opportunities for running test and analysis routines might be more important than studying techniques for generating tests composed of long sequences of actions.

%% file: related.tex
\section{Related Work}
\label{sec:related}

Our study provides an initial characterization of the factors that might cause field failures. The most closely related work includes empirical studies about software faults and techniques to address the problem of testing applications in the field. We discuss both categories below.

\smallskip
\textbf{Empirical Studies}
Most of the studies on software faults and failures focus on the distribution of faults and failures across the components of a system~\cite{Ostrand-TheDistributionOfFaults-ISSTA-2002,Fenton-QuantitativeAnalysisOfFailuresTSE-2000, Runeson-FailureStudy-TSE-2007, GalinacGrbac-TSE-FailureStudy-2013,Shatnawi-2008}.  These and our studies address different goals, but share some hypotheses and observations: they all (i) distinguish between pre-release and post-released  faults, that is, faults detected during and after development, respectively, (ii) provide evidence that often a small subset of the components of a system contains most of the post-release faults, and (iii) indicate that the size of a component is not a good predictor of the post-release fault density~\cite{Ostrand-TheDistributionOfFaults-ISSTA-2002,Fenton-QuantitativeAnalysisOfFailuresTSE-2000, Runeson-FailureStudy-TSE-2007, GalinacGrbac-TSE-FailureStudy-2013}. 

%
The results reported in existing studies might be exploited to improve the design of techniques for predicting failures and locating faults in the field, \CHANGED{for example Wu et al.~\cite{Wu-QuantitativeFailuresAnalysis-ESEM-2008} report a direct dependency between the quality of the testing process and the density of pre- and post-release faults in the source files}. However  they do not contribute in the identification of the factors that may cause and motivate the presence of field failures.

Both Hamill et al.~\cite{Hamill-FaultTypesDetectionSeverity-SQJ-2014} and Fan et al.~\cite{Fan-NuclearFailures-SF-2013} proposed taxonomies of field-faults, with a high-coarse granularity and a focus on nuclear applications, respectively.
Hamill et al. identify a set of field-fault categories at a granularity level that is much higher than ours and provide limited support for analyzing the characteristics of field failures in details. 
%
%
Hamil et al. conclude that coding faults are the major cause of field failures, but do not further analyze the characteristic of the faults, such as the nature of the triggers of the failures, as done in this paper, where we discuss the relation between field failures and the interactions with external resources. 

Fan et al. show that in the nuclear industry software field failures mostly depend on design choices~\cite{Fan-NuclearFailures-SF-2013}. Although the paper does not investigate why these faults have not been revealed at an earlier stage of development, some of the failure causes reported in the paper are consistent with the results that we obtained. For example, the presence of failures caused by incorrect assumptions and unexpected execution conditions are consistent with the failures that we reported under the unknown environment or application conditions categories. The consistency with the results reported by our study increases the confidence on the validity of the results reported in this paper.


\begin{changed}
Grottke et al.~\cite{grottke2010empirical} and Cotroneo et al.~\cite{cotroneo2013fault} study  Bohrbugs and Mandelbugs in safety critical applications and open-source software, respectively.  Both studies focused on variations of amount of Bohrbugs and Mandelbugs during the application life cycle and on their impact on the time to fix a bug. 
Both studies reveal some dependency of Mandelbugs from interaction of the software with field elements. Our results confirm that dependencies of some bugs from environment interactions persist also after deployment. 
Lutz et al.~\cite{lutz2004empirical} classify safety-critical anomalies using Orthogonal Defect Classification, focusing on how to improve safety-critical software development process. 
Our study identifies anomaly types and targets fixes after deployment.
\end{changed}

\smallskip
\textbf{Testing In the Field}
Field testing techniques are techniques that aim to reveal faults that escape in house testing before causing the system to fail in the field.
Field testing has been addressed quite recently with In Vivo Testing~\cite{Murphy-InVivo-ICST-2009}, Skoll~\cite{Skoll2007} and in the context of Web services~\cite{Hielscher2008, DenaroTosi2009,Sammodi2011}. 

In Vivo testing is a technique for identifying faults that are triggered only in specific program states~\cite{Murphy-InVivo-ICST-2009}. The approach targets Java classes and consists of executing a predefined set of test cases while the software is running in the field.  
Test cases are executed within a sandboxed replica of the system, with a parametric frequency and at randomly selected time. 
In Vivo testing can detect faults that can be triggered with a predefined set of inputs provided by the software developers, but cannot detect faults that are triggered with inputs that have not been identified at testing time. Moreover, the strategy does not take into account the factors that demonstrate to play a key role in field failures according to our study, such as the interaction with the field. 

Skoll aims to identify the faults that have not been detected at testing time because of the combinatorial explosion of the configuration options and 
\CHANGED{the characteristics of the environment~\cite{Skoll2007}}. 
Skoll distributes testing tasks across machines of volunteering end-users to extensively explore and test the configuration space. 
Like in-vivo testing, Skoll does not generate test cases nor reuses runtime data to enhance testing activities. Thus, it cannot detect faults that are revealed by inputs not considered by the software engineers who implemented the test cases deployed in the field. 


In the context of Web-service composition, online testing is used to trigger predefined self-adaptation strategies when Web services do not behave as expected, for example because the service specification has been updated or because the service itself is down
~\cite{Hielscher2008, DenaroTosi2009, Sammodi2011}. 
These approaches target a specific class of software systems, and rely on strategies and inputs predefined by the software engineers.

The results reported in this paper call for novel testing and analysis techniques that can be executed in the field to reveal field-intrinsic faults. These techniques must have the ability to test situations that cannot be foreseen at development time and must face the challenge of dealing with field elements, such as local resources and services, without interfering with the user activity.

%% file: conclusion.tex
\section{Conclusion}
\label{sec:conclusion}

\balance

This paper reports the results of an empirical study about the characteristics of field failures, that is, failures observed in production environments.
In details, we introduce the concept field-intrinsic faults as faults inherently hard to detect in-house and more effectively detectable in the field. Field-intrinsic faults are a relevant subset of the more general category of field faults that characterize faults in the field regardless of why they escaped in-house testing. 
We report our findings about the high frequency of field-intrinsic faults in the analyzed bug reports (field-intrinsic faults represent 70\% of the analyzed field faults), obtaining initial evidence that there is a relevant amount of faults that cannot be effectively addressed in-house and should be addressed directly in the field. 
We qualitatively analyze the cases, and identify four main reasons for the presence of field-intrinsic faults: cases impossible to replicate in-house, combinatorial explosion of the cases that should be tested, unknown application or environment conditions.

We investigate the characteristics of these faults to determine the elements that may make them intrinsically hard to detect. We identify the need to reason on the state of the application and the need to interpret the outputs of the application as two key features of techniques designed to reveal these faults.

We are currently continuing the experimental evaluation of reports extracted from the bug repositories of applications in the same domain, to assess the quantitative data reported in this paper, and in different domains, to study the impact of the domains on field failures. 
\CHANGED{reveal}
field-intrinsic faults.